\documentclass{article}

\usepackage{arxiv}

\usepackage[utf8]{inputenc} 
\usepackage[T1]{fontenc}    
\usepackage{hyperref}       
\usepackage{url}            
\usepackage{booktabs}       
\usepackage{amsmath,amssymb,amsfonts}      
\usepackage{nicefrac}       
\usepackage{microtype}      
\usepackage{lipsum}         
\usepackage{graphicx}
\usepackage{natbib}
\usepackage{doi}

\usepackage{amsmath,amssymb,amsfonts}
\usepackage{amsthm}
\usepackage{cleveref}       

\usepackage{mathtools}
\usepackage{mathpartir}
\usepackage{stmaryrd}
\usepackage{caption}
\usepackage{subcaption}

\theoremstyle{definition}
\newtheorem{definition}{Definition}[section]
\newtheorem{example}{Example}[section]
\theoremstyle{plain}
\newtheorem{theorem}{Theorem}[section]

\theoremstyle{remark}

\usepackage{proof}

\usepackage{fancyvrb}
\usepackage{listings}
\usepackage{caption}
\usepackage{algorithmic}
\usepackage{graphicx}
\usepackage{textcomp}
\usepackage{xcolor}
\def\BibTeX{{\rm B\kern-.05em{\sc i\kern-.025em b}\kern-.08em
    T\kern-.1667em\lower.7ex\hbox{E}\kern-.125emX}}

\newcommand{\braces}[1]{{\{}#1{\}}}

\title{Model-Driven Security Analysis of Self-Sovereign Identity Systems}

\date{}

\newif\ifuniqueAffiliation
\uniqueAffiliationtrue

\ifuniqueAffiliation 
\author{ \href{https://orcid.org/0000-0002-6996-9333}{\includegraphics[scale=0.06]{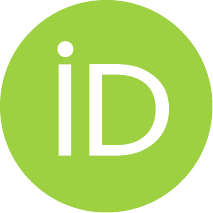}\hspace{1mm}Yepeng~Ding}\thanks{\url{https://yepengding.github.io/}} \\
	The University of Tokyo \\
 	Tokyo, Japan\\
	\texttt{yepengding@g.ecc.u-tokyo.ac.jp} \\
	\And
	\href{https://orcid.org/0000-0002-2891-3835}{\includegraphics[scale=0.06]{orcid.pdf}\hspace{1mm}Hiroyuki Sato} \\
	The University of Tokyo\\
	Tokyo, Japan\\
	\texttt{schuko@satolab.itc.u-tokyo.ac.jp} \\
}
\else
\usepackage{authblk}

\newbox{\orcid}\sbox{\orcid}{\includegraphics[scale=0.06]{orcid.pdf}} 
\author[1]{%
	\href{https://orcid.org/0000-0000-0000-0000}{\usebox{\orcid}\hspace{1mm}David S.~Hippocampus\thanks{\texttt{hippo@cs.cranberry-lemon.edu}}}%
}
\author[1,2]{%
	\href{https://orcid.org/0000-0000-0000-0000}{\usebox{\orcid}\hspace{1mm}Elias D.~Striatum\thanks{\texttt{stariate@ee.mount-sheikh.edu}}}%
}
\affil[1]{Department of Computer Science, Cranberry-Lemon University, Pittsburgh, PA 15213}
\affil[2]{Department of Electrical Engineering, Mount-Sheikh University, Santa Narimana, Levand}
\fi

\begin{document}
\maketitle

\begin{abstract}
Best practices of self-sovereign identity (SSI) are being intensively explored in academia and industry. Reusable solutions obtained from best practices are generalized as architectural patterns for systematic analysis and design reference, which significantly boosts productivity and increases the dependability of future implementations. For security-sensitive projects, architects make architectural decisions with careful consideration of security issues and solutions based on formal analysis and experiment results. In this paper, we propose a model-driven security analysis framework for analyzing architectural patterns of SSI systems with respect to a threat model built on our investigation of real-world security concerns. Our framework mechanizes a modeling language to formalize patterns and threats with security properties in temporal logic and automatically generates programs for verification via model checking. Besides, we present typical vulnerable patterns verified by SecureSSI, a standalone integrated development environment, integrating commonly used pattern and attacker models to practicalize our framework.
\end{abstract}


\section{Introduction}
Self-sovereign identity (SSI) is an emerging digital identity model that aims to empower physical entities with full control over their digital identities. Unlike traditional identity models, SSI promotes decentralized approaches in identification and authentication mechanisms to overcome challenges in centralized architectures, such as regulatory compliance, privacy protection, availability, and dependability, which is gaining momentum worldwide as it potentially neutralizes the centralized digital identity control by the oligopoly. With the advancement of cloud computing and blockchain technology, various SSI systems have been implemented based on the W3C standard, catering to a wide range of scenarios, such as Internet of Things \cite{fedrecheski_self-sovereign_2020}, healthcare \cite{houtan_survey_2020}, finance \cite{de_cristo_self-sovereign_2021}, and data aggregation \cite{ding_leveraging_2022}. Meanwhile, exploring the best practice of SSI systems becomes a promising research direction in academia and sustaining engineering work in industry. Based on the theoretical studies and practical experiences, architectural patterns are extracted from the best practice as reusable solutions to increase productivity and lower business risks.

Architectural patterns of SSI systems can be categorized based on SSI components: issuer agent (IA), holder agent (HA), verifier agent (VA), and verifiable data registry (VDR) \cite{ding_self-sovereign_2022}. We extract commonly used patterns from academic studies \cite{tobin_inevitable_2016,liu_design_2020,houtan_survey_2020,naik_uport_2020,fedrecheski_self-sovereign_2020,de_cristo_self-sovereign_2021,soltani_survey_2021,ding_self-sovereign_2022} and our investigation of 11 real-world SSI systems, including seven commercial products (e.g., Microsoft Entra\footnote{https://docs.microsoft.com/en-us/entra/}), three open-source libraries (e.g., Veramo\footnote{https://veramo.io/}), and two standalone platforms (e.g., SMART Health Cards\footnote{https://smarthealth.cards/}). For instance, decentralized VDR is a coarse-grained architectural pattern that narrows the tech stack of VDR implementations to decentralized technologies. Fine-grained architectural patterns derived from the decentralized VDR further determine the implementation details, such as public-blockchain-based and private-blockchain-based patterns. The self-management pattern, a coarse-grained architectural pattern of a HA's verifiable credential (VC) manager, can be refined as the cold wallet pattern using the external device to manage VCs in local environments. These architectural patterns guide the development of SSI systems with insights into core factors, such as cost, security, and user experience.

Security stands as a crucial dimension in evaluating architectural patterns. Architectural patterns tailored for security \cite{yoder_architectural_1997} have been widely adopted to bolster the strength of systems. IN the context of SSI systems, security takes center stage along with controllability and portability \cite{tobin_inevitable_2016} during the design and implementation process. However, existing studies on the security analysis of SSI systems remain insufficient in theory and practice. While most studies focus on solving security issues related to identity management \cite{kuperberg_blockchain-based_2019,fedrecheski_self-sovereign_2020,houtan_survey_2020,de_cristo_self-sovereign_2021} by introducing SSI, they solely emphasize the importance of distributed ledger technologies (DLTs) for enhancing SSI system security. Only a few studies examine security concerns of SSI systems from practical perspectives, such as identity proving \cite{othman_horcrux_2018}, network security \cite{stokkink_truly_2021}, and risk evaluation \cite{naik_evaluation_2022}. However, these perspectives alone are not adequate for the security analysis of rapidly evolving architectural patterns \cite{liu_design_2020,ding_self-sovereign_2022}. Moreover, model-based methods that specify and verify architectural security properties for security analysis are rarely studied.

Formal models use mathematical languages to describe structures, behaviors, and communications of systems rigorously. Besides, systems are specified with properties that are proved via formal verification techniques, such as model checking \cite{clarke_model_1997,baier_principles_2008} and theorem proving \cite{davis_machine_1962,fitting_first-order_2012}. These specification and verification techniques collectively fall under the umbrella of formal methods, playing a pivotal role in the security analysis of various types of systems \cite{ritchey_using_2000,celik_soteria_2018,wang_formal_2018}. Formal methods enable the detection of vulnerabilities by searching proofs that violate the formulated properties and provide a way to find attack paths by searching counterexamples to the given properties. Furthermore, experiments can be designed and conducted for empirical analysis based on the proofs and counterexamples.

Motivated by filling the void of systematic security analysis of SSI systems, we propose a model-driven security analysis framework to analyze the security properties of architectural patterns of SSI systems regarding a threat model. Our framework 1) adopts a modeling language to formally specify architectural patterns and threats; 2) formally verifies properties via model checking with automatically generated NuSMV programs; 3) forges skeleton programs conforming verified properties in Java for redevelopment or repairment. Besides, our framework preserves extensibility and scalability by modularizing coarse-grained patterns and enables module refinement for fine-grained patterns so that verified properties are reusable at multiple abstraction levels. Our contributions are summarized as follows.

\begin{itemize}
\item
We identify security concerns of eight commonly used architectural patterns based on our analysis of surveyed academic studies and 11 real-world SSI systems. We further formulate a threat model normalizing security concerns.

\item
We propose a novel model-driven security analysis framework that formalizes architectural patterns and threats and formally verifies security properties encoding threat consequences defined in our threat model.

\item
We present typical vulnerable patterns verified by SecureSSI, a standalone integrated development environment (IDE) implementing our proposed framework.

\end{itemize}

\section{Related Work}
\label{sec:related_work}
SSI has been extensively studied in academia \cite{houtan_survey_2020,fedrecheski_self-sovereign_2020,de_cristo_self-sovereign_2021,ding_leveraging_2022} and practiced in industry \cite{tobin_inevitable_2016,naik_uport_2020}. As a relatively new concept, best practices are still under exploration. Meanwhile, to extend the applicability of SSI and facilitate the development of SSI systems, design and architectural patterns \cite{liu_design_2020,ding_self-sovereign_2022} are extracted from implementations.

The work \cite{liu_design_2020} presents a collection of design patterns for blockchain-based SSI from the perspective of key management, DID management, and credential design. Although the security and privacy of the presented patterns are discussed, no formal and experimental analyses are given. In work \cite{ding_self-sovereign_2022}, SSI systems are encapsulated as services to provide out-of-the-box functionalities. Architectural patterns for SSI services are organized based on four components: endorsement component, wallet component, verification component, and data component, with respect to the IA, HA, VA, and VDR. However, the pattern evaluation of security aspects is not systematic and lacks experiments for demonstration.

We find many works such as \cite{stokkink_deployment_2018,houtan_survey_2020,kondova_self-sovereign_2020} advocate the adoption of DLTs to enhance the security of SSI systems, especially integrity and availability. A comprehensive study is presented in work \cite{ferdous_search_2019} to show the significance of adopting blockchain technology with a formal model. In \cite{ferdous_search_2019}, the authors present a mathematically defined digital identity model as the fundamental of modeling SSI based on the first-order logic. Furthermore, the authors discuss the role of blockchains in empowering the SSI with concrete protocols. 

Besides, some studies \cite{othman_horcrux_2018,stokkink_truly_2021,naik_evaluation_2022} propose solutions to specific security issues in SSI systems. SSI does not have an identity-proving mechanism to link DIDs to the represented physical identities. The authors of \cite{othman_horcrux_2018} introduce biometric-based authentication (IEEE Standard for Biometric Open Protocol) in an SSI system to enable identity proving. This method effectively prevents masquerade attacks by detecting the biometric traits of holders. Network privacy issues potentially lead to unauthorized disclosure of identity information in SSI systems. Motivated by addressing privacy issues at the network level, the authors of \cite{stokkink_truly_2021} propose a solution named TCID to provide network-level anonymity with acceptable latency. Besides, the authors of \cite{stokkink_truly_2021} and \cite{soltani_survey_2021} point out that zero-knowledge proof techniques are also promising to solve privacy issues in SSI systems. Motivated by evaluating potential attacks on SSI systems, the work \cite{naik_evaluation_2022} proposes an approach based on an attack tree model and risk matrix model that identifies three potential attacks: faking identity, identity theft, and distributed denial of service attacks. This work also provides insights into the risk management of developing SSI systems.

\section{Self-Sovereign Identity}
\label{sec:ssi}
Self-sovereign identity (SSI) is usually enabled by decentralized identifiers (DIDs)\footnote{https://www.w3.org/TR/did-core/} and verifiable credentials (VCs)\footnote{https://en.wikipedia.org/wiki/Verifiable\_credentials}. Each physical entity is identified by a unique DID determined by the controller, which decouples identification from centralized identity providers (IdPs) and certificate authorities (CAs). IdPs and CAs are also assigned with DIDs and are probably indistinguishable from common users. Identity information is included in VCs and issued from an entity called issuer to another entity called holder that preserves VCs. Holders prove their identities by presenting VCs, or verifiable presentations (VPs) composed of VCs issued from a set of issuers to entities called verifiers. Verifiers can verify the validity of VCs and VPs based on the proofs and issuer identities obtained from a verifiable data registry (VDR). Therefore, entities are generally classified into three roles: issuer, holder, and verifier.

From a technical view, we regard an SSI system as a comprehensive system containing four components: IA (issuer agent), HA (holder agent), VA (verifier agent), and VDR (verifiable data registry). An agent acts on behalf of an entity to communicate with other components. All three types of agents interact with VDR to register and resolve DIDs. Additionally, IAs are controlled by issuers to issue VCs for holders. Holders control HAs to request and store VCs, as well as present VCs to verifiers. VAs are controlled by verifiers to collect and verify VCs to authenticate holders.


\subsection{Architectural Patterns}
\label{sec:architectural_patterns}
We categorize architectural patterns of SSI systems into four types: IA patterns, HA patterns, VA patterns, and VDR patterns, which correspond to the SSI components. We extract and summarize commonly used coarse-grained architectural patterns of SSI systems in Table~\ref{tab:coarse_patterns} based on the studies \cite{tobin_inevitable_2016,liu_design_2020,houtan_survey_2020,naik_uport_2020,fedrecheski_self-sovereign_2020,de_cristo_self-sovereign_2021,soltani_survey_2021,ding_self-sovereign_2022} and our investigated 11 real-world SSI systems.

\begin{table}[htbp]
\caption{Coarse-grained architectural patterns of SSI systems.}
\begin{center}
\begin{tabular}{|l|l|}
\hline
\textbf{Component Name} & \textbf{Pattern Name}\\
\hline
Issuer Agent & Single-Key Pattern (SK), Multi-Key Pattern (MK) \\
\hline
Holder Agent & Self-Management Pattern (SM), Vendor-Management Pattern (VM) \\
\hline
Verifier Agent & Dependent Pattern (DT), Independent Pattern (IT) \\
\hline
Verifiable Data Registry & Centralized Pattern (CD), Decentralized Pattern (DD) \\
\hline
\end{tabular}
\label{tab:coarse_patterns}
\end{center}
\end{table}

\subsection{Threat Model}
\label{sec:threat_model}
We formulate a threat model for the security concerns about architectural patterns based on RFC 4949 \cite{shirey_internet_2007}. Our threat model is composed of a passive threat model and an active threat model. Each threat model contains a set of attacks composed of a sequence of threat actions. In this section, we show attacks and threat consequences with intuitive examples.

\subsubsection{Passive Threat Model}
The passive threat model contains attacks monitoring component behaviors of SSI systems and eavesdropping on communications between components. These attacks usually result in threat consequences that threaten confidentiality, such as unauthorized disclosure of internal states of components and message contents in communications. In most cases, passive attacks are performed in silence and are hard to detect. Besides, attackers can utilize the information sniffed and inferred from the success of passive attacks to perform active attacks.

For instance, a vulnerable HA based on the \textit{Vendor-Management Pattern} follows an insecure communication protocol that can be intercepted by man-in-the-middle (MitM) attacks, potentially resulting in unauthorized disclosure of VCs. In the worst case, leaked VCs are reused by attackers to impersonate the victim holder to deceive verifiers via attacks in the active threat model.

\subsubsection{Active Threat Model}
Attacks in the active threat model manipulate data streams and forge false streams to launch attacks such as replay attacks, masquerade attacks, data tampering, and denial of service (DoS). The threat consequences from these attacks can be deception, disruption, and usurpation, which damage integrity and availability.

A typical example is a vulnerable VDR based on the \textit{Centralized Pattern}, which has a poorly implemented centralized service providing APIs for identity and credential registration and verification in a weakly protected network. Attackers can tamper with the database by exploiting vulnerabilities like CVE-2021-44228\footnote{https://cve.mitre.org/cgi-bin/cvename.cgi?name=cve-2021-44228} and further launch masquerade attacks, which results in deception and usurpation. The weakly protected network without effective traffic migration techniques may encounter distributed denial of service (DDoS) attacks that causes disruption.

We also consider internal and external collusion as a representative attack in the active threat model because we do not presume a trust chain such as holders trust verifiers protect the privacy of presented VCs and VPs, verifiers trust issuers issue correct identity information, and issuers trust holders properly use issued VCs. Internal collusion happens when an administrator of an IA or a VA is paid a hefty monetary reward and becomes fearless in impairing the normal functioning of the agent. External collusion happens when two roles collude with each other and threaten another role. For example, a holder colludes with an issuer to sign a falsified VC used to deceive a verifier.

\section{Model-Driven Analysis}
In this section, we present our model-driven analysis framework to formalize architectural patterns of SSI systems and malicious entities. We introduce our modeling method and illustrate pattern and threat modeling based on the Seniz language through a running example.

\subsection{Modeling Method}
\label{sec:modeling_method}

We elaborate a modeling method that separately models structures and behaviors for patterns (Section~\ref{sec:pattern_modeling}) and threats (Section~\ref{sec:threat_modeling}).

Our modeling method revolves around a labeled transition system (LTS) defined in Definition~\ref{def:lts} based on the variant of the transition system initially formulated in \cite{ding_formalism-driven_2022-1} to formalize architectural patterns and malicious entities.

\begin{definition}[Labeled Transition System]
\label{def:lts}
Over set \textit{Var} of typed state variables, a labeled transition system $\mathfrak{T}$ is a tuple
\begin{equation*}
    \mathfrak{T} \triangleq \langle S, A, {\hookrightarrow}, I, g_0, P, \mathcal{L} \rangle
\end{equation*}

where
\begin{itemize}
    \item $S = \llbracket \textit{Var} \rrbracket$ is a set of states,
    \item $A$ is a set of actions,
    \item ${\hookrightarrow} \subseteq S \times \| \textit{Var} \| \times A \times S$ is a conditional transition relation,
    \item $I \subseteq S$ is a set of initial states,
    \item $g_0 \in \| \textit{Var} \|$ is the initial condition,
    \item $P$ is a set of atomic propositions, and
    \item $\mathcal{L}: S \mapsto \wp(P)$ is a labeling function.
\end{itemize}

$\llbracket \textit{Var} \rrbracket$ represents the set of evaluations of state variables \textit{Var}, which determines the state space $S$. Actions are carried upon transitions to emit observable effects. We use the notation $s \xhookrightarrow{g \triangleright a} s'$ as shorthand for $(s, g, a, s') \in {\hookrightarrow}$ where $g \in \| \textit{Var} \| $ is a transition guard composed of a set of propositional formulae over \textit{Var}. In this manner, the execution of action $a$ is only triggered when $\mathcal{V} \models g$ for transition $s \xhookrightarrow{g \triangleright a} s'$ where $\mathcal{V} \in \llbracket \textit{Var} \rrbracket$. An atomic proposition is an indecomposable proposition defined in propositional logic, which can be observed by behaviors. Particularly, we consider $S$, $A$, and $P$ as finite sets.
\end{definition}

Given a $\mathfrak{T}$ over set $P$ of atomic propositions, we can specify its properties in temporal logic, including linear temporal logic (LTL) \cite{sistla_complexity_1985} and computation tree logic (CTL) \cite{emerson_decision_1982}.

\subsubsection{Structure Modeling}
Structures of architectural patterns consist of state variables with types, including Boolean, integer, string, and enumeration. State variables are assignable with typed values matching their declared types. According to Definition~\ref{def:lts}, state $s \in S$ is determined by an assignment $\mathcal{V} \in \llbracket \textit{Var} \rrbracket$ of all state variables defined in a labeled transition system, where $\mathcal{V}$ is also called a configuration or an evaluation of $s$.

However, manually setting all configurations is impractical for modeling complex systems. Particularly, many state variables are constants or frozen variables that remain unchanged throughout the lifespan. It is inconvenient to set these variables for each configuration repeatedly. Besides, a pattern generally has a collection of components to be modeled. Managing configurations across all components globally is challenging.

To overcome the difficulties of modeling complex patterns, we refine methods introduced in \cite{ding_formalism-driven_2022} based on the system graph defined in Definition~\ref{def:sys_graph}.

\begin{definition}[System Graph]
\label{def:sys_graph}
A system graph $\mathfrak{S}$ without terminal state declarators over set \textit{Var} of typed state variables is a tuple
\begin{center}
$\mathfrak{S} \triangleq \langle D, \mathcal{N}, A, {\hookrightarrow}, i, g_0, C, \mathcal{L} \rangle$
\end{center}
where
\begin{itemize}
    \item $D=N \times \llbracket \widehat{\textit{Var}} \rrbracket, \widehat{\textit{Var}} \subseteq \textit{Var}$ is a set of state declarators with names in $N$,
    \item $\mathcal{N}: D \mapsto \wp(\llbracket \textit{Var} \rrbracket)$ is a naming function,
    \item $A$ is a set of actions,
    \item ${\hookrightarrow} \subseteq D \times \| \textit{Var} \| \times A \times D$ is the conditional transition relation,
    \item $i \in D$ is the initial state declarator,
    \item $g_0 \in \| \textit{Var} \|$ is the initial guard,
    \item $C \supseteq \| \textit{Var} \|$ is a set of clauses, and
    \item $\mathcal{L}: \llbracket \textit{Var} \rrbracket \mapsto \wp(C)$ is a labeling function.
\end{itemize}

Here, the uniqueness rule below also applies.
\[
\forall \langle n, \widehat{\mathcal{V}} \rangle \in D: (\nexists \langle n', \widehat{\mathcal{V}}' \rangle \in D: \widehat{\mathcal{V}} = \widehat{\mathcal{V}}' \land n \neq n').
\]

The clause set $C$ is formulated by atomic propositions and logical connectives. A set of clauses are related to any configuration via the labeling function $\mathcal{L}$.

\end{definition}

Based on Definition~\ref{def:sys_graph}, we can use system graphs to identify a subset of state variables with state declarators and infer the evaluation of unidentified state variables automatically, instead of manually setting all configurations. Besides, with the refinement and modularization support \cite{alur_alternating_1998,pasareanu_concrete_2005,baier_principles_2008,ding_formalism-driven_2022}, we can disassemble a pattern into subsystems and modules and evolve the model from a primitive structure, which can significantly simplify the modeling process.

\subsubsection{Behavior Modeling}
Behaviors of architecture patterns are defined by conditional transition relations and specified by temporal logic. For similar reasons, we also use system graphs to describe behaviors.

According to Definition~\ref{def:sys_graph}, a conditional transition relation is composed of a source state declarator, a transition guard, an action, and a target state declarator. Both source and target state declarators are formulated while modeling structures. A behavior happens if one of the configurations identified by the source state declarator satisfies the transition guard and is observed by action effects emitted from the action.

\paragraph{Action Elimination}
\label{sec:act_elim}
As a modeling process, the formal analysis does not require the implementation of functionalities, i.e., overriding and observing action effects are unnecessary. Therefore, some actions can be eliminated to optimize the behavior modeling.

We present this approach with an example. Consider a HA model $\mathfrak{S}^h$ with transition relation ${\to_1} \in {\hookrightarrow}$ defined as
$
\textit{WaitVCRes} \xhookrightarrow{(\textit{isCodeSuccess}) \triangleright (\textit{ParseVCRes})} \textit{HaveVCRes}.
$

Here, we have identified the configurations of state declarators \textit{WaitVCRes} and \textit{HaveVCRes} according to Definition~\ref{def:sys_graph}. Action \textit{ParseVCRes} is declared to emit the effects of parsing the request for logging purposes only. Hence, state variables remain unchanged before and after the execution of \textit{ParseVCRes}. Consequently, $\mathfrak{S}^h$ will go from \textit{WaitVCRes} to \textit{HaveVCRes} if $\mathcal{V}_{WaitVCRes} \models \textit{isCodeSuccess}$ with or without carrying action \textit{ParseVCRes}. Therefore, we can simply remove \textit{ParseVCRes} and use $\epsilon$ to mark ${\to}_1$ as an epsilon transition.

Notably, an action cannot be eliminated if it has effects on state variables. For example, transition relation ${\to_2} \in {\hookrightarrow}$ of $\mathfrak{S}^h$ is defined as $\textit{ReadyForVCReq} \xhookrightarrow{(\textit{isParamCorrect}) \triangleright (\textit{SendVCReq})} \textit{WaitVCRes}$.

It is reasonable to identify the change of state variables associated with the request in action \textit{SendVCReq}, which requires that $\mathfrak{S}^h$ must go through an intermediate state generated by the action effect of \textit{SendVCReq} to reach \textit{WaitVCRes}. Therefore, we cannot directly eliminate action \textit{SendVCReq}. Otherwise, the unintended behavior of $\mathfrak{S}^h$ will occur. We can circumvent this by delegating state variable changes in the effect of \textit{SendVCReq} to \textit{ReadyForVCReq} or \textit{WaitVCRes}.

\paragraph{Behavior Specification}
Specifying behaviors with temporal logic is another aspect of behavior modeling. Our method enables directly formulating temporal properties with LTL and CTL over system graphs to enhance the readability of behavior specifications. Furthermore, our method interprets LTL and CTL formulas over the states and paths of a labeled transition system defined in Definition~\ref{def:lts}. This interpretability facilitates model checking of formulated properties and enables concrete counterexample generation over labeled transition systems, of which the proof is shown as follows.

\begin{theorem}[Interpretability]
\label{th:interp}
For any system graph $\mathfrak{S}$ without terminal state declarators, there is an interpretation based on a labeled transition system defined in Definition~\ref{def:lts}.

\begin{proof}
Giving a system graph $\langle D, \mathcal{N}, A^s, {\hookrightarrow}^s, i, g_0^s, C, \mathcal{L}^s \rangle$ over set \textit{Var} of typed state variables, we can construct a labeled transition system $\langle S, A^t, {\hookrightarrow}^t, I, g_0^t, P, \mathcal{L}^t \rangle$ where
\begin{itemize}
    \item $S = \bigcup\limits_{d \in D} \mathcal{N}(d)$,
    \item $A_t = A_s$,
    \item ${\hookrightarrow}^t = \{ s \xhookrightarrow{g \triangleright a} s' \mid s \in \mathcal{N}(d), s' \in \mathcal{N}(d'), (d, a, g, d') \in {\hookrightarrow}^s \}$,
    \item $I = \mathcal{N}(i)$,
    \item $g_0^t = g_0^s$,
    \item $P = \bigcup\limits_{c \in C} \mathcal{D}(c)$, and
    \item $\mathcal{L}^t(s) = \bigcup\limits_{c \in \mathcal{L}^s(s)} \mathcal{D}(c)$ with $\mathcal{D}$ as a decomposition function that extracts all atomic propositions composing the given clause.
\end{itemize}
\end{proof}
\end{theorem}

With Theorem~\ref{th:interp}, we can model structures based on system graphs and specify behaviors with properties formulated in LTL and CTL without additional techniques for model checking.

Continuing the example in Section~\ref{sec:act_elim}, we can specify the behavior of $\mathfrak{S}^h$: \textit{whenever VC request $\textit{req}$ is ready, then VC response $\text{res}$ is received} in LTL as $\Box (\textit{req.isReady} \to \Diamond (\textit{res.isOk}))$.

We also specify the behavior: \textit{It is inevitable that the HA invariantly has a DID} in CTL as .
$\forall \Diamond \forall \Box (did \neq \varnothing)$.

Notably, the second behavior specified in CTL has no equivalent LTL formula due to the incomparable expressiveness of LTL and CTL.

\subsection{Pattern Modeling}
\label{sec:pattern_modeling}

To construct system graphs and formulate temporal properties, we use a subset of the Seniz language \cite{ding_formalism-driven_2022}. The Seniz language provides a collection of syntactic sugar to facilitate the modeling process. Besides, the compiler can generate Graphviz programs for visualization, Promela \cite{holzmann_model_1997} programs for LTL property verification, and NuSMV \cite{cimatti_nusmv_2002} programs for CTL property verification. Furthermore, minimal implementations of the architectural patterns can be developed based on the generated Java skeleton programs. 



\subsubsection{Workflow}
\begin{figure}[htbp]
\centerline{\includegraphics[scale=0.44]{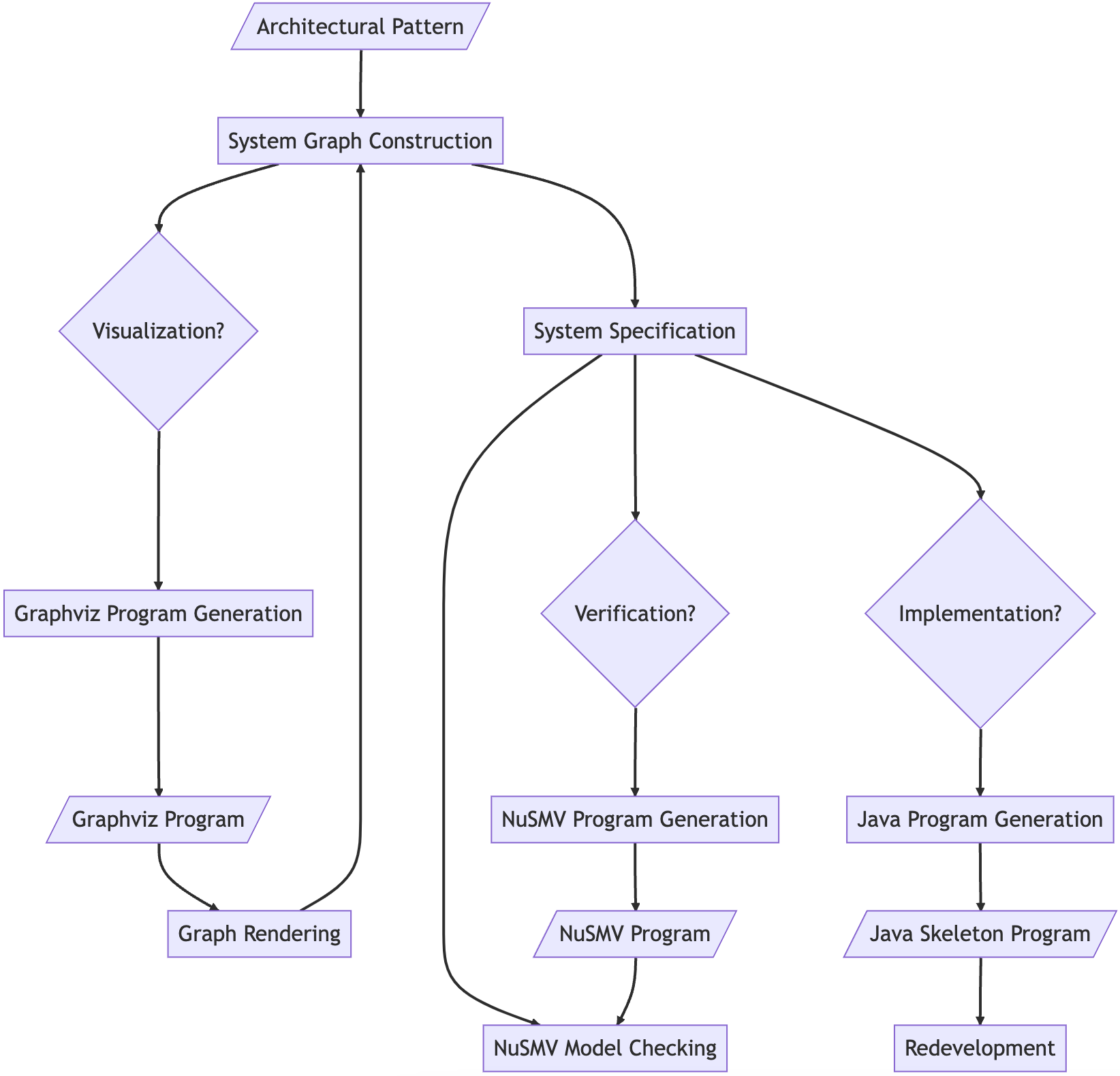}}
\caption{Overview of the workflow of modeling architectural patterns.}
\label{fig:workflow}
\end{figure}

As shown in Figure~\ref{fig:workflow}, we first construct a system graph $\mathfrak{S}$ for an architectural pattern in a visualizable way with the refinement and modularization support \cite{alur_alternating_1998,pasareanu_concrete_2005} provided by Seniz \cite{ding_formalism-driven_2022}. Then we define a finite set of system specifications $Q = \braces{\Phi_1, \Phi_2, \dots}$. Thus, the verification problem amounts to proving that $\mathfrak{S} \models Q$ via model checking supported by NuSMV. Finally, we generate Java skeleton programs that implement $\textit{Arch}$ via Seniz.

\subsubsection{Modeling Outline}
Architectural patterns rely on different environments like network and concurrency environments, of which their common functionalities are generalized and encapsulated into corresponding modules. The \textit{network module} exposes a static record to simulate network communications, including requests and responses at different network layers. Similar to the \textit{network module}, the \textit{concurrency module} models internal interactions within a component, including message passing and memory sharing in processes and threads.


Each component is independently abstracted as a system graph. Generally, a component contains a set of constants describing immutable attributes, such as system identifier, key pair, and interacting target set, which are declared as system arguments, a syntactic sugar supported by the Seniz language for declaring local variables passed from control systems and remaining unchanged throughout the lifespan. Besides, variables in environment modules are interpreted as global shared variables and introduced into component models.

Therefore, we can construct a protocol context containing an environment model and a set of component models by mapping environment models into component models and declaring interleaving relations for component models to simulate asynchronous execution.

\begin{example}[Protocol Modeling]
\label{eg:protocol_model}
Let us consider a simplified VC uploading protocol of the \textit{Vendor-Management Pattern} where HA uploads a VC to the vendor management service.

We show a snippet of the HA system graph in Listing~\ref{lst:ha}. Here, we omit unrelated logic by rewriting them with a single state variable \textit{state}. We also simplify the encryption processes by encoding them into guards of transition relations that require corresponding keys to receive the message. The constants such as owned and known public keys (\textit{pkH}, \textit{pkV}) are declared as system arguments. To avoid naturally formed terminal states, we append an epsilon transition $\textit{DONE} \xhookrightarrow{\top \triangleright \epsilon} \textit{DONE}$ to the system graph.

\begin{lstlisting}[caption={Program snippet of the HA system graph in Example~\ref{eg:protocol_model}.}, language=Java,mathescape=true, numbers=left, label={lst:ha}, basicstyle=\footnotesize]
...
system Holder(pkH :: string, vcH :: string, vcHash :: string) over HolderVars with Environment {

    init IDLE -> REQ_PK_V -> @sendGetPKvReq sendGetPKvReq() WAIT_PK_V
    ...
    HAVE_VC_CONF -> DONE -> DONE
    ...
    @sendGetPKvReq = {
        reqGetPKvPayload: pkH
    }
    ...
    HAVE_PK_V = {
        state: "HAVE_PK_V",
        pkV: resGetPKvPayloadContent
    }
    ...
    prop receivedConf {
        resPostVCPayloadKey = pkH & resPostVCPayloadContent = vcHash
    }
    
    prop isDone {
        state = "DONE"
    }
    ...
}

varset HolderVars {
    pkV :: string,
    state :: string
}
\end{lstlisting}

We model the main control system in Listing~\ref{lst:main} to initialize all environment state variables as \textit{"NONE"} and declare an asynchronous relation between an instantiated HA system graph and a vendor system graph. We also declare a CTL formula to specify the behavior: \textit{both HA and vendor system graphs will inevitably finish the execution of the protocol and go into the \textit{DONE} state declarator}, i.e., $\forall \Diamond (h.\textit{isDone} \land v.\textit{isDone})$.

\begin{lstlisting}[caption={Program snippet of the main control system in Example~\ref{eg:protocol_model}.}, language=Java,mathescape=true, numbers=left, label={lst:main}, basicstyle=\footnotesize]
...
main control system Main() over Environment {

    init {
        reqGetPKvPayload : "NONE",
        resGetPKvPayloadKey : "NONE",
        resGetPKvPayloadContent : "NONE",
        ...
    }

    async Holder("PK_H", "VC_H", "VC_H_HASH") as h, Vendor("PK_V") as v
    
    ctl AF (h.isDone and v.isDone)
}
\end{lstlisting}

Based on the generated NuSMV program (222 lines in total), we can obtain the report shown in Figure~\ref{fig:vm_protocol_report} from the NuSMV 2.6.0.

\begin{figure}[htbp]
\centerline{\includegraphics[scale=0.6]{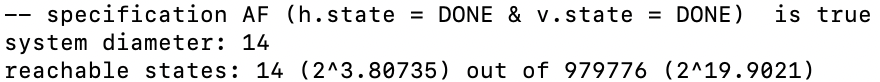}}
\caption{NuSMV running report in Example~\ref{eg:protocol_model}.}
\label{fig:vm_protocol_report}
\end{figure}

\end{example}

\subsection{Threat Modeling}
\label{sec:threat_modeling}
We abstract attackers as system graphs encoding threat actions to find possible attack paths leading to threat consequences shown in Section~\ref{sec:threat_model}. We specify behaviors about a wholistic model consisting of an attacker model and a set of pattern models, so that our framework can prove the threat-free properties by searching counterexamples implying effective attack paths via model checking techniques. If no counterexample is found, we can conclude that the system satisfies the formulated threat-free properties. Otherwise, we can obtain an attack path that violates a property and reproduce it in practice.

\subsubsection{Workflow}
Similar to the workflow of modeling patterns, constructing an attacker model is a process of assigning powers to attackers by formulating threat actions that manipulate a set of state variables, including global shared variables in environment modules and pattern state variables.

We formulate Example~\ref{eg:threat_model} to exhibit the general steps of proving the existence of a passive attack in the protocol illustrated in Example~\ref{eg:protocol_model}.

\begin{example}[Attacker Modeling]
\label{eg:threat_model}
For the protocol formalized in Example~\ref{eg:protocol_model}, we consider the possibility of the MitM attack, a typical passive attack, to check whether an attack path leading to the unauthorized disclosure of the VC exists. The overview of the attack is depicted in Figure~\ref{fig:vm_protocol_mitm}. The middle man tries to intercept the request of $\textit{pk}_v$ and return a malicious public key $\textit{pk}_m$, which is possible to make the HA misread $\textit{pk}_m$ as the public key of the vendor and make the vendor regard the middle man as a legitimate HA.

\begin{figure}[htbp]
\centerline{\includegraphics[scale=0.44]{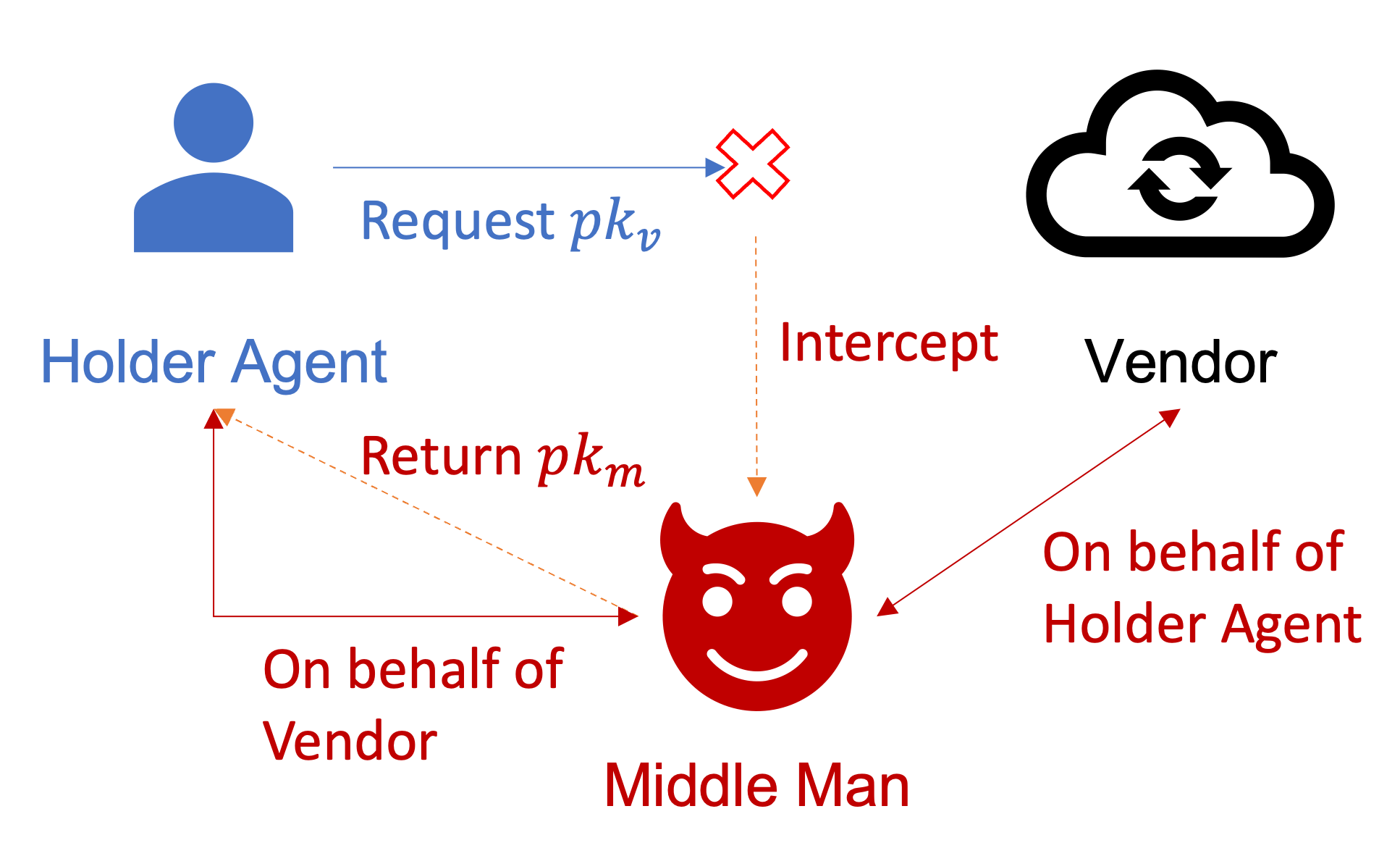}}
\caption{The overview of the MitM attack in Example~\ref{eg:threat_model}.}
\label{fig:vm_protocol_mitm}
\end{figure}

Based on the same modeling methods introduced in Section~\ref{sec:modeling_method}, we formalize the attacker in Listing~\ref{lst:mitm}.

\begin{lstlisting}[caption={Program snippet of the attacker in the middle in Example~\ref{eg:threat_model}.}, language=Java,mathescape=true, numbers=left, label={lst:mitm}, basicstyle=\footnotesize]
...
system MiddleMan(pkM :: string) over MiddleManVars with Environment {
    init IDLE -> IDLE [reqGetPKvPayload] -> parseInterceptedGetPKvPayload() HAVE_PK_H -> sendManipulatedGetPKvReq() WAIT_PK_V
    WAIT_PK_V -> WAIT_PK_V [resGetPKvPayloadKey = pkM] -> parseGetPKvPayload() HAVE_PK_V -> sendMaliciousPKv() RES_PK_M -> WAIT_VC
    WAIT_VC -> WAIT_VC [reqPostVCPayloadKey = pkM] -> parseInterceptedPostVCPayload() HAVE_VC
    HAVE_VC -> HAVE_VC [resPostVCPayloadKey = pkM] -> parseVCConfPayload() HAVE_VC_CONF -> sendMaliciousVCRes() RES_UPLOAD_VC
    RES_UPLOAD_VC -> DONE -> DONE
    ...
}

varset MiddleManVars {
    pkH :: string,
    pkV :: string,
    vc :: string,
    state :: string
}
\end{lstlisting}

We append an instantiated attacker $m$ into the declared asynchronous relation in the main control system in Listing~\ref{lst:main}. As a passive attack, the MitM attack should not perturb the normal functionalities of the protocol. Hence, we preserve the declared CLT property to ensure both the HA and vendor system will normally finish the execution of the protocol and go into the \textit{DONE} state declarator. Besides, we declare a new CTL formulate for the behavior: \textit{there does not exist a path where the middle man eventually has the VC of the HA}, i.e., $\nexists \Diamond (m.vc = h.vcH)$. The result of executing the generated NuSMV program (353 lines in total) is shown in Figure~\ref{fig:vm_protocol_mitm_report}.

\begin{figure}[htbp]
\centerline{\includegraphics[scale=0.44]{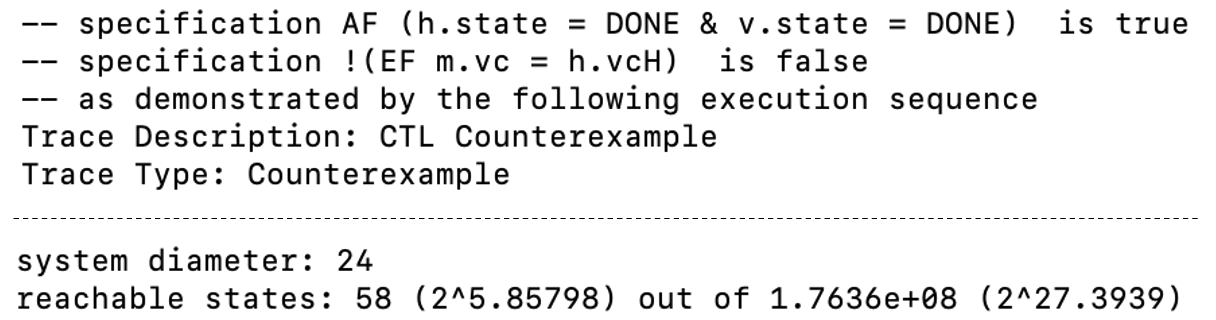}}
\caption{NuSMV running report in Example~\ref{eg:threat_model}.}
\label{fig:vm_protocol_mitm_report}
\end{figure}

From the result, we can obtain that the first property holds while the second one fails, meaning that the MitM attack succeeds without being detected, and the model checker produces an attack path.

\end{example}

\subsubsection{Modeling Outline}
In this section, we introduce the modeling outline for four representative attacks: masquerade, MitM, collusion, and DoS, which cover most types of passive and active threat consequences in our threat model in Section~\ref{sec:threat_model}.

\paragraph{Masquerade}
We abstract a network environment visible to the attacker and parsable if the communication is unencoded or the attacker has knowledge about decoding (e.g., key and nouce). For masquerade attackers, we mechanize possible threat actions that redirect the traffic into the malicious component of the attacker model by searching and exploiting the unencoded communications. With the intercepted knowledge, we encode the impersonating logic into the attacker model and formulate properties to check the existence of attack paths.

\paragraph{Man-in-the-middle}
MitM attackers are modeled in a similar way to masquerade attackers. The main difference is that MitM attacks have additional steps to bridge normal communications between two components. As an attack in the passive threat model, MitM attacks sniff the information while remaining undetected. Hence, the formulated properties include a property implying undetectability.

\paragraph{Collusion}
A collusion attack is modeled based on the nondeterministic transition relations generated by malicious fragments of components. A malicious fragment can access the bribery channel formulated as an exceptional environment model that establishes the exchange of insider information. The properties of the success of a collusion attack are formulated as ensuring the existence of an attack path leading to a state where the malicious fragment knows some unauthorized information. 

\paragraph{Denial-of-service}
We encode nondeterministic transition relations into the DoS attack model, potentially making a component enter its \textit{Unavailable} state. This attack model exploits internal mechanisms of components like task priority without fairness assumptions. We formulate safety properties that check components never go into their \textit{Unavailable} states.

%

\section{SecureSSI}

\begin{table*}[htbp]
\caption{Typical vulnerable patterns verified by SecureSSI.}
\begin{center}
\resizebox{\textwidth}{!}{\begin{tabular}{|l|c|l|l|}
\hline
\textbf{Target Patterns} & \textbf{Violated Property Example} & \textbf{Description} \\
\hline
($\mathfrak{M}$) \{ MK \} & $\forall \Diamond i.\textit{DONE} \land \forall \Box (\bigwedge\limits_{a \in I} a.\textit{sk} \notin m.\textit{disclose})$ & Private key sniffing from communications between a collector and administrators. \\

($\mathfrak{M}$) \{ VM \} & $\forall \Diamond (h.\textit{DONE} \land v.\textit{DONE}) \land \forall \Box (h.\textit{attr} \notin m.\textit{disclose})$ & Unauthorized disclosure of holder attributes from communications within HAs. \\

($\mathfrak{M}$) \{ SK, VM \} & $\forall \Diamond (h.\textit{DONE} \land i.\textit{DONE}) \land \forall \Box (h.\textit{vcH} \notin m.\textit{disclose})$ & VC sniffing from communications between IAs and HAs and within HAs. \\

($\mathfrak{C}$) \{ SK \} & $\forall \Box (m.\textit{VC\_ISSUED} \implies (h.\textit{attr} \succ i.\textit{vcH}))$ & Internal collusion of IAs manipulating VC issuing. \\

($\mathfrak{C}$) \{ IT \} & $\forall \Box ((h.\textit{attr} \succ m.\textit{vcH})) \implies \forall \Diamond m.\textit{VC\_VERIFIED}$ & Internal collusion of VAs manipulating VC verification results. \\

($\mathfrak{C}$) \{ SK, SM \} & $\forall \Box (h.\textit{attr} \succ v.\textit{vcH})$ & External collusion of HAs forging VCs through corrupted IAs. \\

($\mathfrak{Q}$) \{ SK, CD \} & $\forall \Box (m.\textit{pkM} \notin i.\textit{vcH} \land m.\textit{attr} \notin i.\textit{vcH})$ & Impersonating a legitimate HA to deceive an IA. \\

($\mathfrak{Q}$) \{ SM, CD \} & $\forall \Box (m.\textit{pkM} \notin i.\textit{vcH} \land h.\textit{attr} \notin m.\textit{disclose})$ & Impersonating a legitimate IA to deceive a HA. \\

($\mathfrak{Q}$) \{ IT, DD \} & $\nexists \Diamond (m.\textit{pkM} \in i.\textit{vcH} \implies \forall \Diamond v.\textit{VC\_VERIFIED})$ & Impersonating a legitimate HA to deceive a VA. \\

($\mathfrak{D}$) \{ MK \} & $\forall \Box (h.\textit{REQ\_SENT} \implies \forall \Diamond i.\textit{VC\_ISSUED})$ & Unavailable collector. \\

($\mathfrak{D}$) \{ VM \} & $\forall \Box (h.\textit{REQ\_SENT} \implies \forall \Diamond h.\textit{vcH} \neq \emptyset)$ & Unavailable vendor. \\

($\mathfrak{D}$) \{ DT \} & $\forall \Box (v.\textit{REQ\_SENT} \implies \forall \Diamond v.\textit{vcH} \neq \emptyset)$ & Unavailable third-party system. \\

\hline
\end{tabular}}
\label{tab:typical_violations}
\end{center}
\end{table*}

We developed SecureSSI, a standalone IDE in Java and TypeScript, to practicalize the proposed model-driven security analysis framework. SecureSSI integrates reusable models of architectural patterns in Table~\ref{tab:coarse_patterns} and two environment modules (network module and concurrency module). Besides, we implement four preset attacker models: MitM ($\mathfrak{M}$), collusion ($\mathfrak{C}$), masquerade ($\mathfrak{Q}$), and DoS ($\mathfrak{D}$) in SecureSSI.

In Table~\ref{tab:typical_violations}, we present typical vulnerable patterns verified by SecureSSI concerning specific attacks with examples of violated properties. The first column show the target patterns with the attacker model in parentheses and the architectural patterns in brackets. The latter two columns give concrete examples of threat consequences in CTL with descriptions. Here, we use phrases with capital letters to abbreviate entity states. We use $\textit{attr}$ and $\textit{disclose}$ to denote the ground truth of an entity embedded as credentials in VCs and disclosed information storage. For brevity, we also introduce function $\succ$ to denote the conformity between the ground truth and a VC, evaluated to $\top$ if the ground truth conforms to the data in a VC.

\section{Conclusion}
In this paper, we formulated a threat model based on our investigated security concerns of architectural patterns in academic studies and real-world SSI systems. Motivated by providing insights for systematic security analysis of SSI systems, we proposed a model-driven security analysis framework that formalizes architectural patterns and threats to SSI systems with respect to the threat model. Additionally, we present typical vulnerable patterns verified by SecureSSI, a standalone IDE integrating our proposed framework and equipped with built-in pattern and attacker models for practical security analysis of real-world SSI systems.

\bibliographystyle{unsrtnat}
\bibliography{references}  






\end{document}